\documentclass[12pt]{article}
\usepackage{latexsym}
\newcommand{\tr}[1]{\,{\rm tr}\,#1\,}

\begin{document}
\title{ \vspace{0.5cm} Higgs mechanism as a collective effect due to extra dimension.}
\author{A.A.Slavnov.\\Steklov Mathematical Institute
\\Gubkina st. 8, 119991 Moscow, \\Moscow State University}
\maketitle

\begin{abstract}
A systematic analysis of the unitary electroweak model described by
the higher derivative Lagrangian depending on extra dimension
\cite{Sl1} is presented.
\end{abstract}

\section{Introduction}

The essential part of renormalizable models of weak and
electromagnetic interactions, like Weinberg-Salam model \cite{We},
\cite{Sa}, or more advanced models describing neutrino
oscillations is the spontaneous symmetry breaking via interaction
with the spin zero Higgs meson \cite{Hi}, \cite{Ki}.

The Higgs meson interaction is described by the Lagrangian of the
form
\begin{eqnarray}
L=|\partial_{\mu} \varphi+ig \frac{\tau^a}{2}A_{\mu}^a \varphi-
\frac{ig_1}{2}B_{\mu}\varphi|^2+\nonumber\\
G[(\bar{L} \varphi)R+ \bar{R}( \varphi^+L)]+ \ldots +
\frac{m^2}{2}(\varphi^+\varphi)-\lambda^2(\varphi^+ \varphi)^2
\label{1}
\end{eqnarray}
Here $A_{\mu}, B_{\mu}$ are the $SU(2)$ and $U(1)$ gauge fields,
$\varphi$ is the complex doublet, $L,R$ denote the chiral lepton
$SU(2)$ multiplets and $ \ldots$ stand for the similar terms
corresponding to quarks.

The minimum of the Higgs potential is achieved at some nonzero
nonsymmetric value of $\varphi$ and the perturbation theory near
the stable minimum may be constructed in terms of the shifted
scalar fields having the following real components
\begin{equation}
\varphi_1= \frac{iB_1+B_2}{\sqrt{2}}; \quad \varphi_2= \frac{
\sqrt{2}m_1}{g}+ \frac{1}{\sqrt{2}}(\sigma-iB_3) \label{2}
\end{equation}
where $m_1$ is the mass which the Yang-Mills field acquires via
Higgs mechanism.

The gauge invariance of the model allows to choose the gauge
condition $B_a=0$, corresponding to the manifestly unitary gauge,
whereas another choice, for example $\partial_{\mu}A_{\mu}=0$,
provides a power-counting renormalizable perturbation theory
\cite{Ho}.

Gauge invariance allows to eliminate from the spectrum all
components $B_a$, but the field $\sigma$ survives and must be
observed experimentally. The mass of this field is $m_2=2 \sqrt{2}
\lambda m_1g^{-1}$ and at first sight it seems that choosing the
constant $\lambda$ big enough one may shift this mass to an
unobservable region. However this mass enters also the self
interaction of Higgs mesons, which in the unitary gauge acquires a
form:
\begin{equation}
L_H=- \frac{gm_2^2}{4m_1} \sigma^3- \frac{g^2m_2^2}{32m_1^2}
\sigma^4 \label{3}
\end{equation}
One sees that in the limit $m_2 \rightarrow \infty$ this
interaction blows up violating the renormalizability of the
theory.

Existing experimental data impose stringent limits on the mass of
the Standard Model Higgs meson. These limits may be changed by
different modifications of the Weinberg-Salam model. In particular
the models with additional scalar gauge singlet mesons were
discussed  (see e.g. \cite{CaDo}, \cite{HiBi}, \cite{ChiGo},
\cite{Bj}). These modifications indeed change predictions concerning
masses and other characteristics of scalar mesons, but at the same
time they introduce considerable ambiguity in the choice of the
Lagrangian, describing the interaction of the singlet spin zero
particles.

An interesting idea was put forward by N.Krasnikov \cite{Kr}, who
considered the limiting case when the number of scalar singlets is
infinite. In this case one can avoid the appearance of one particle
poles in the channels corresponding to the exchange by a spin zero
particle and in this sense to get rid off fundamental scalar mesons.

A common deficiency of all above mentioned models is a lack of a
guiding principle of choosing a particular form of the singlet field
interactions.

The form of the interaction of additional gauge singlet fields may
be fixed by some symmetry of a corresponding Lagrangian. This
approach was adopted in our paper \cite{Sl1} and is studied in more
details in the present paper. Our model is similar to the model
proposed in the paper \cite{Kr}, and in the same sense does not
require the existence of  fundamental spin zero bosons. We interpret
the spontaneous symmetry breaking responsible for the mass
generation of intermediate vector mesons as a collective effect
related to the dependence of the singlet scalar fields on the
additional coordinate. The Yang-Mills fields and the Higgs doublet
fields are living on the four dimensional brane in the five
dimensional space, whereas the gauge singlet scalar fields and their
masses depend also on the fifth coordinate. The form of the singlet
field interaction is fixed by requirement of invariance of the
Lagrangian with respect to some supersymmetry transformations. This
symmetry arises in a natural way in the higher derivative model
\cite{Sl1} and will be discussed below.

\section{Spontaneous symmetry breaking via extra dimension.}

To avoid trivial complications in this section we consider the
gauge group $SU(2)$ and only vector meson - scalar sector.

Following the paper \cite{Sl1} we assume that the gauge fields
$A_{\mu}$, as well as the fermions and the spin zero complex
doublet $\varphi$ are living on the four dimensional "brane" and
transform under the gauge transformations as in the Weinberg-Salam
model. In addition we introduce the gauge singlet neutral scalar
field, which depends on the extra coordinate $\lambda: X(x,
\lambda)$.

The dynamics of the model is described by the higher derivative
Lagrangian, which has a form:
\begin{eqnarray}
L=L_{YM}+(D_{\mu}\varphi)^+(D_{\mu}\varphi)+\nonumber\\
\frac{g}{2m_1}\partial_{\mu}(\varphi^+ \varphi) \int_{-
\pi/\kappa}^{\pi/\kappa} \partial_{\mu}X(x, \lambda) d \lambda+
\frac{1}{2} \int_{- \pi/\kappa}^{\pi/\kappa}
\partial_{\mu}X(x, \lambda) d \lambda \int_{-\pi/\kappa}^{\pi/\kappa}
\partial_{\mu}X(x, \lambda) d \lambda+\nonumber\\
\frac{\pi}{\kappa} \int_{-\pi/\kappa}^{\pi/\kappa} \Box X(x,
\lambda) \Box X(x, \lambda)a^{-2}(\lambda) d \lambda-
\frac{\pi}{\kappa} \int_{-\pi/\kappa}^{\pi/\kappa}
\partial_{\mu}X(x, \lambda) \partial_{\mu}X(x, \lambda)
d\lambda+\nonumber\\
\int_{- \pi/\kappa}^{\pi/\kappa} \partial_{\mu} \bar{c}(x,
\lambda) \partial_{\mu}c(x, \lambda)d \lambda
 \label{4}
\end{eqnarray}
where $\varphi$ is again parameterized as in eq.(\ref{2}). The
fields $\bar{c},c$ are anticommuting scalar ghosts, singlet with
respect to gauge transformations. We assume that the extra
dimension is compact $ -\pi/\kappa \leq \lambda \leq \pi/\kappa$.
The function $a^{-2}( \lambda)$, which determines the masses of
the $X$-fields also may depend on $\lambda$.

To quantize the model we start with the discretized version,
introducing the lattice in the fifth dimension. We take $ \frac{2
\pi}{\kappa}=Nb$, where $b$ is the lattice spacing, $\lambda_i=bi,
\quad X_i=X(\lambda_i), \quad c_i=c(\lambda_i), \quad
\bar{c}_i=\bar{c}(\lambda_i)$. After rescaling the fields $
\sqrt{b}c_i \rightarrow c_i, \sqrt{b} \bar{c}_i \rightarrow
\bar{c}_i, bX_i \rightarrow X_i$ the discretized Lagrangian may be
written in the form:
\begin{eqnarray}
L=L_{YM}+(D_{\mu}\varphi)^+(D_{\mu}\varphi)+ \frac{g}{2m_1}
\partial_{\mu}(\sum_{i=1-N/2}^{N/2} X_i) \partial_{\mu}(\varphi^+
\varphi)+\nonumber\\
\frac{1}{2}( \sum_{i=1-N/2}^{N/2} \partial_{\mu}X_i)^2-
\frac{N}{2} \sum_{i=1-N/2}^{N/2}
\partial_{\mu}X_i
\partial_{\mu}X_i+\nonumber\\
\frac{N}{2} \sum_{i-1-N/2}^{N/2}a_i^{-2} \Box X_i \Box X_i+
\sum_{i=1-N/2}^{N/2}
\partial_{\mu} \bar{c}_i \partial_{\mu}c_i \label{7}
\end{eqnarray}
The Lagrangian (\ref{7}) includes higher derivatives and one may
expect the appearance of negative norm states. They indeed appear,
but due to invariance of this Lagrangian under supersymmetry
transformations which will be described below the negative norm
states decouple and the theory is unitary in the nonnegative norm
subspace \cite{Sl2}, \cite{Sl3}.

To quantize the model we shall use the Ostrogradsky canonical
formalism \cite{Sl4}, \cite{Sl3}. First of all we diagonalize the
quadratic form in the Lagrangian (\ref{7}) by the shift
\begin{equation}
\sigma(x) \rightarrow \sigma(x)- \sum_{i=1-N/2}^{N/2} X_i
\label{9}
\end{equation}
After such a shift the Lagrangian takes a form:
\begin{eqnarray}
L= \tilde{L}(A_{\mu},B, \tilde{\sigma})+ \frac{1}{2}
\partial_{\mu} \sigma \partial_{\mu} \sigma - \frac{N}{2}
\sum_{i=1-N/2}^{N/2}
\partial_{\mu}X_i \partial_{\mu}X_i+\nonumber\\
\frac{N}{2} \sum_{i=1-N/2}^{N/2}a_i^{-2} \Box X_i \Box X_i+
\sum_{i=1-N/2}^{N/2}
\partial_{\mu} \bar{c}_i
\partial_{\mu}c_i+ \frac{g}{4m_1} \partial_{\mu}(B^2+ \tilde{\sigma}^2)
\sum_{i=1-N/2}^{N/2} \partial_{\mu}X_i \label{10}
\end{eqnarray}
Here $ \tilde{L}+ \frac{1}{2} \partial_{\mu} \sigma \partial_{\mu}
\sigma$ is the usual Lagrangian for the massive Yang-Mills field
interacting with the scalar fields $B_a, \sigma$, in which
$\sigma$ in the interaction is replaced by
\begin{equation}
 \tilde{\sigma}= \sigma- \sum_{i=1-N/2}^{N/2} X_i
 \label{7a}
 \end{equation}

This Lagrangian is obviously invariant with respect to the gauge
transformations
\begin{eqnarray}
\delta A_{\mu}^a=(D_{\mu} \varepsilon)^a \nonumber\\
\delta \sigma=- \frac{g}{2}(B_a \varepsilon^a)\nonumber\\
\delta B_a=-m_1 \varepsilon_a- \frac{g}{2} \varepsilon_{abc}B^b
\varepsilon^c- \frac{g}{2} \sigma \varepsilon_a+ \frac{g}{2}
\sum_{i=1-N/2}^{N/2} X_i \varepsilon_a \label{11}
\end{eqnarray}
Using this invariance we impose the gauge condition $B_a=0$.
Quantization of the fields $A_{\mu}^a, \sigma, \bar{c}_i, c_i$ is
performed in a standard way.

The Lagrangian for $X_i$ field includes higher time derivatives
and to quantize it we introduce the variables $Q_i^1=X_i, \quad
Q_i^2= \dot{X}_i$ and conjugated momenta
\begin{equation}
P_1^i=-N \partial_0 X^i-a_i^{-2}N \partial_0 \Box X_i+
\tilde{f}(A_{\mu}, \sigma,X^i); \quad P_2^i=a_i^{-2}N \Box X^i
\label{12}
\end{equation}
Here $ \tilde{f}$ denotes the terms of order $O(g)$.

In terms of these variables the Fourier transform of the
Hamiltonian looks as follows
\begin{eqnarray}
\tilde{H}(\textbf{k})= \sum_i P_1^iQ_2^i+ \sum_i P_2^i
\dot{Q}_2^i-L
= \sum_i\{P_1^iQ_2^i- \textbf{k}^2P_2^iQ_1^i+\nonumber\\
 \frac{(P_2^ia_i)^2}{2N}+ \frac{N}{2}(Q_2^i)^2+ \frac{N}{2} \textbf{k}^2(Q_1^i)^2
 +\frac{1}{2}(P_{\sigma})^2+\nonumber\\
\frac{1}{2}\textbf{k}^2 \sigma^2+P_c^iP_{\bar{c}}^i+ \textbf{k}^2
\bar{c}_ic_i+ \tilde{H}_1(A_{\mu}, \tilde{\sigma},X) \label{13}
\end{eqnarray}
In this equation $\tilde{H}_1$ denotes the Hamiltonian of the free
massive vector field and the interaction terms.

The spectrum of the free Hamiltonian includes a spin zero massless
one-particle state generated by the field $ \sigma$, as well as
$N$ spin zero massless and $N$ spin zero massive one-particle
states with the masses $a_i$, generated by the fields $X_i$. The
corresponding creation and annihilation operators are
\begin{equation}
\sigma^{\pm}(\textbf{k})= \frac{1}{\sqrt{2 \omega}}( \omega \sigma
\mp ip_{\sigma}); \quad [  \sigma^-(\textbf{k}),
\sigma^+(\textbf{k}')]= \delta(\textbf{k}-\textbf{k}'); \quad
\omega= \sqrt{\textbf{k}^2} \label{14}
\end{equation}
\begin{equation}
a^{\pm}_i(\textbf{k})= \frac{1}{\sqrt{2 \omega N}}(P_1^i \mp iN
\omega Q_1^i \mp i \omega P_2^i); \quad
[a_i^-(\textbf{k}),a_i^+(\textbf{k}')]=-
\delta(\textbf{k}-\textbf{k}') \label{15}
\end{equation}
\begin{equation}
b_i^{\pm}(\textbf{k})= \frac{1}{\sqrt{2 \omega N}}(P_1^i+NQ_2^i
\mp i \omega_iP_2^i); \quad
[b^-_i(\textbf{k}),b^+_j(\textbf{k}')]= \delta_{ij}
\delta(\textbf{k}-\textbf{k}'); \quad
\omega_i=\sqrt{\textbf{k}^2+a_i^2} \label{16}
\end{equation}

The operators $a^+$ create massless states with nonpositive norm.
However due to invariance of the Lagrangian (\ref{10}) with
respect to the supersymmetry transformations
\begin{eqnarray}
\delta X_i(x)=c_i(x) \varrho; \quad\delta \sigma(x)
= \sum_{i=1-N/2}^{N/2}c_i(x) \varrho;\nonumber\\
 \delta \bar{c}_i(x)=[NX_i(x) -\sigma(x) -
Na_i^{-2} \Box X_i(x) - \frac{g}{4m_1}(B(x)^2+
\tilde{\sigma}(x)^2)] \varrho, \label{8}
\end{eqnarray}
where $\varrho$ is an anticommuting parameter, there exists a
conserved operator $Q$, which allows to select the nonnegative
norm states by imposing the condition
\begin{equation}
Q|\Phi>=0 \label{18}
\end{equation}
The asymptotic operator $Q_0$ has a form
\begin{eqnarray}
Q_0= \int d^3k \sum_i(NQ_1^i \partial_0c^i+P_1^ic_i+P_2^i
\partial_0c^i-P_{\sigma}c_i+ \sigma \partial_0c_i)=\nonumber\\
\int d^3k \sum_i[c_i^+( \sqrt{N}a_i^-- \sigma^-)+( \sqrt{N}a_i^+-
\sigma^+)c_i^-] \label{19}
\end{eqnarray}
The solution of the eq.(\ref{18}) for asymptotic states has a form
\begin{equation}
|\Phi>=[1+\sum_n b_n \prod_{1\leq j \leq
n}(\sigma^+(\textbf{k}_j)- \frac{1}{\sqrt{N}}
\sum_{i=1-N/2}^{N/2}a_i^+(\textbf{k}_j))]|\Phi>_{A,b_i} \label{20}
\end{equation}
where the vectors $|\Phi>_{A,b_i}$ describe the states of the
massive vector field and the scalar particles with the masses
$a_i$. The vectors (\ref{20}) have nonnegative norm and this
subspace being factorized with respect to the null vector space
coincides with the physical one.

It is worth to mention that for $N \neq 1$ the conserved operator
$Q$ is not nilpotent. However as in our case the ghost fields
$c_i, \bar{c}_i$ are noninteracting, the condition (\ref{18})
guarantees the positivity of the physical states norm.

It completes the quantization of the model in the "unitary"gauge
$B_a=0$. In this gauge the theory is not power counting
renormalizable as the vector field propagator
\begin{equation}
\tilde{D}^{ab}_{\mu \nu}(k)= \delta^{ab} \frac{g^{\mu
\nu}-k^{\mu}k^{\nu}m_1^{-2}}{k^2-m_1^2} \label{21}
\end{equation}
does not decrease at infinity. However using the gauge invariance
of the Lagrangian (\ref{10}) one can pass in a standard way to
some renormalizable gauge, i.e. the Lorentz gauge
$\partial_{\mu}A_{\mu}=0$. In this gauge the $S$-matrix is given
by the path integral
\begin{equation}
S= \int \exp \{i\int Ldx\} \delta(\partial_{\mu}A_{\mu}) \det[M]
dA_{\mu}dB^ad \sigma dX_i \label{22}
\end{equation}
where $L$ is the gauge invariant Lagrangian (\ref{10}), $\det[M]$
is the Faddeev-Popov determinant and proper boundary conditions in
the path integral are assumed. We omitted the ghost fields $
\bar{c}_i, c_i$ as being free they do not influence the result.

The propagators, corresponding to the integral (\ref{22}) are
\begin{eqnarray}
\tilde{D}^{ab}_{\mu \nu}(k)= \delta^{ab} \frac{g^{\mu
\nu}-k^{\mu}k^{\nu}k^{-2}}{k^2-m_1^2} \nonumber\\
\tilde{D}^{ab}_{FP}(k)=\delta^{ab} \frac{1}{k^2}\nonumber\\
\tilde{D}_{\sigma}(k)= \frac{1}{k^2}\nonumber\\
\tilde{D}^{ab}_{B}(k)= \delta^{ab} \frac{1}{k^2}\nonumber\\
\tilde{D}_X^{ij}(k) = \delta^{ij}\frac{a_i^2} {k^2(k^2-a_i^2)N}
=\frac{\delta^{ij}}{N} [ \frac{1}{k^2-a_i^2}- \frac{1}{k^2}]
\label{23}
\end{eqnarray}
All the propagators except for $\tilde{D}_X(k)$ decrease at $k
\rightarrow \infty$ as $k^{-2}$, and the ultraviolet asymptotic of
the propagator $\tilde{D}_X(k)$ is $k^{-4}$. Obviously the model
is renormalizable.

The fields $X$ which may produce nonpositive norm states enter the
interaction only in the combinations
\begin{equation}
\tilde{\sigma}= \sigma- \sum_iX_i, \quad \tilde{X}_i=\Box X_i
\label{24}
\end{equation}
In the $ \tilde{\sigma}$ propagators the contribution of the field
$\sigma$ compensates exactly the contribution of zero mass
components of $X_i$ and only positive norm components with the
masses $a_i$ are propagating. The propagators of $ \tilde{X}_i$
also do not have zero mass poles. Neither of these combinations
generates negative norm asymptotic states in accordance with the
previous proof of the unitarity in the physical subspace.

The probability of the creation of a spin zero state with the mass
$a_i$ is suppressed by the factor $N^{-1}$ and in the limit $N
\rightarrow \infty$ vanishes. However as the number of $X_i$
fields in the intermediate states in this limit goes to infinity,
their total contribution is finite and describes some collective
effect. This contribution depends on the particular form of the
function $a(i)$. If the dependence of $a$ on $i$ is such that for
$N \rightarrow \infty$ the series
\begin{equation}
\sum_{i=1-N/2}^{N/2} \frac{1}{k^2-a_i^2} \label{24a}
\end{equation}
is convergent, then in this limit
\begin{equation}
\sum_{i,j=1-N/2}^{N/2} \tilde{D}_X^{ij}(k) \sim \frac{1}{k^2}.
\label{24b}
\end{equation}
As the interaction of $X$ fields includes the second derivative,
one obtains a nonrenormalizable theory.

The other extreme case corresponds to $a_i\equiv a$. In this case
\begin{equation}
\sum_{i,j}\tilde{D}_X^{ij}(k)= \frac{a^2}{k^2(k^2-a^2)}
\label{24c}
\end{equation}
and the propagator of the $\tilde{\sigma}$-field is
\begin{equation}
\tilde{D}_{\tilde{\sigma}}(k)= \frac{1}{k^2-a^2} \label{25}
\end{equation}
It is nothing but the usual Higgs model. All massive spin zero
states have the same quantum numbers and are undistinguishable.

In a general case for large momenta $k^2 \gg a_i^2$, the propagator
$ \tilde{D}_{\tilde{ \sigma}}(k)$ coincides with the usual Higgs
meson propagator $ \tilde{D}_{\tilde{\sigma}} \sim k^{-2}$, but for
$k^2 \simeq a_i^2$ the predictions of our model may be very
different of the Higgs-Kibble model. For any finite $N$ the
Lagrangian (\ref{10}) describes a model with $N$ neutral spin zero
particles with the masses $a_i$, which is unitary in the positive
norm space and renormalizable . For $N$ big enough the individual
scalar states become unobservable, but their presence in the
intermediate states produces some collective effects which may be
checked experimentally.

The model described above may be also described by a Lagrangian
without higher derivative terms introducing a gauge invariant
interaction of the singlet fields with the Higgs meson as it was
done in the papers cited in the introduction. In distinction of
these papers in our approach the spin zero Higgs field, described by
the complex doublet field $\varphi $ is eliminated completely from
the spectrum of observables and all physical spin zero excitations
are assosiated with the gauge singlets. Even more important is the
invariance of the Lagrangian (\ref{7}) with respect to the
supersymmetry transformation (\ref{8}), which together with the
requirement of renormalizability fixes the form of the interaction
of the singlet fields. In particular it forbides the vertices of the
type $X^3$ or $ \varphi^+ \varphi X^2$.

To come back to the theory with compact continuous extra
dimension, described by the Lagrangian (\ref{4}), one has to
consider the limit $N \rightarrow \infty, b \rightarrow 0, Nb=2
\pi \kappa^{-1}$. This theory will be discussed in the next
section.

\section{Continuous extra dimension.}

We shall consider the model with the continuous extra dimension as
a limiting case of the discrete model (\ref{7}), whose
quantization was discussed in the previous section.

Let us introduce the lattice spacing $b$ by rescaling the fields
$X_i \rightarrow bX_i$. In the limit $N \rightarrow \infty, Nb=2
\pi \kappa^{-1}$ the $S$-matrix will take the form eq.(\ref{22}),
where the Lagrangian is equal to
\begin{eqnarray}
L= \tilde{L}(A_{\mu},B^a, \tilde{\sigma})+ \frac{1}{2}
\partial_{\mu} \sigma \partial_{\mu} \sigma+
\frac{g}{4m_1} \partial_{\mu}(B^2+ \tilde{\sigma}^2) \int_{-
\pi/\kappa}^{\pi/\kappa} d \lambda \partial_{\mu}X(x,
\lambda)-\nonumber\\
\frac{\pi}{\kappa} \int_{-\pi/\kappa}^{\pi/\kappa}d \lambda
\partial_{\mu}X(x, \lambda) \partial_{\mu}X(x, \lambda)+
\frac{\pi}{\kappa} \int_{-\pi/\kappa}^{\pi/\kappa}d \lambda
a^{-2}( \lambda) \Box X(x, \lambda) \Box X(x, \lambda) \label{26}
\end{eqnarray}
with
\begin{equation}
\tilde{\sigma}= \sigma- \int_{-\pi/\kappa}^{\pi/\kappa}d \lambda
X(x, \lambda) \label{27}
\end{equation}
This Lagrangian is obtained from eq.(\ref{4}) by substituting the
explicit expression for $\varphi$ in terms of $B,\sigma$, making
the shift (\ref{27}) and omitting the free ghost fields
$\bar{c}(x, \lambda),c(x,\lambda)$.

The fields $A_{\mu}, B_a, \sigma$ and the Faddeev-Popov ghosts do
not depend on the extra dimension, so their propagators have the
standard form. The $X$-field propagator is easily calculated. It
is equal to
\begin{equation}
\tilde{D}_X(k, \lambda, \mu)= \delta(\lambda-\mu)
\frac{\kappa}{2\pi} \frac{a^2(\lambda)}{k^2(k^2-a^2(\lambda))}
\label{28}
\end{equation}
As before the interaction terms depend only on the combinations
\begin{equation}
\tilde{\sigma}(x)= \sigma(x)- \int_{-\pi/\kappa}^{\pi/\kappa}X(x,
\lambda)d \lambda; \quad
\tilde{X}(x)=\int_{-\pi/\kappa}^{\pi/\kappa} d \lambda \Box X(x,
\lambda) \label{29}
\end{equation}
The corresponding propagators
\begin{eqnarray}
\tilde{D}_{\tilde{\sigma}}(k)= \frac{1}{k^2}+
\int_{-\pi/\kappa}^{\pi/\kappa} d \lambda d \mu
\frac{\kappa}{2\pi}[-\frac{1}{k^2}+ \frac{1}{k^2-a^2(\lambda)}]
\delta(\lambda-\mu)=\nonumber\\
\frac{\kappa}{2\pi} \int_{-\pi/\kappa}^{\pi/\kappa} d \lambda
\frac{1}{k^2-a^2(\lambda)} \label{30}
\end{eqnarray}
\begin{equation}
\tilde{D}_{ \tilde{X}, \tilde{X}}= \int_{-\pi/\kappa}^{\pi/\kappa}
d \lambda \frac{\kappa}{2 \pi}
\frac{k^2a^2(\lambda)}{k^2-a^2(\lambda)} \label{31}
\end{equation}
\begin{equation}
\tilde{D}_{\tilde{X},
\tilde{\sigma}}(k)=\int_{-\pi/\kappa}^{\pi/\kappa}d \lambda
\frac{\kappa}{2 \pi} \frac{a^2(\lambda)}{k^2-a^2(\lambda)}
\label{31a}
\end{equation}
have no one-particle poles, and in this sense in our model there are
no fundamental spin zero particles. The spin zero excitations in
intermediate states produce some collective effects, described by
the propagators (\ref{30}, \ref{31}, \ref{31a}). For a general $a(
\lambda)$ in the limit $|k| \gg \max[a]$ the $ \tilde{
\sigma}$-field propagator coincides with the Higgs meson propagator.
If $a( \lambda)$ is a constant $a( \lambda)=a$, the $\tilde{\sigma}$
propagator coincides with the Higgs meson propagator $
\tilde{D}_H=(k^2-a^2)^{-1}$ for any $k$, and as in the discrete case
the model reduces to the usual Higgs-Kibble model.

An explicit form of the $\tilde{\sigma}$ propagator depends on the
functions $a(\lambda)$, which should be chosen to fit the
experiment. For example, if
\begin{equation}
a^2(\lambda)= M^2- \lambda^2, \quad M^2> \frac{\pi^2}{\kappa^2}
\label{34}
\end{equation}
then
\begin{equation} \tilde{D}_{\tilde{\sigma}}(k)=-\frac{\kappa}{2 \pi \sqrt{M^2-k^2}}
 \ln( \frac{\sqrt{M^2-k^2}-
\pi/\kappa}{\sqrt{M^2- k^2}+ \pi/\kappa}) \label{35}
\end{equation}
Instead of one particle poles we have branch points describing a
collective excitation.

Another example was considered in \cite{Sl1}. If
\begin{equation} a^2(\lambda)= \lambda+
\frac{\pi}{\kappa}+M^2, \quad \lambda<0; \quad a^2(\lambda)=-
\lambda+ \frac{\pi}{\kappa}+M^2, \quad \lambda>0 \label{32}
\end{equation}
then
\begin{equation}
\tilde{D}_{\tilde{\sigma}}(k)=- \frac{\kappa}{\pi} \ln(1-
\frac{\pi}{\kappa (k^2-M^2)}) \label{33}
\end{equation}
As before the one particle poles are replaced by the branch
points, and for $k^2 \gg \frac{\pi^2}{\kappa^2}$
\begin{equation}
\tilde{D}_{\tilde{\sigma}}(k)\simeq k^{-2} \label{35a}
\end{equation}

Choosing properly the function $a(\lambda)$ one can model a
different behaviour of the propagators
$\tilde{D}_{\tilde{\sigma}}, \tilde{D}_{\tilde{X}, \tilde{X}},
\tilde{D}_{\tilde{X}, \tilde{\sigma}}$.

\section{Application to electroweak interactions.}

The $SU(2) \times U(1)$ models of electroweak interactions are
readily generalized to include five dimensional scalar fields. The
gauge meson - fermion sector may be chosen the same as in the
models with the usual Higgs mesons. In the Weinberg-Salam model it
is described by the Lagrangian
\begin{eqnarray}
L_{GF}= \frac{1}{8} \tr[F_{\mu \nu}F_{\mu \nu}]- \frac{1}{4}G_{\mu
\nu}G_{\mu \nu}+ \bar{L}i \gamma_{\mu}(\partial_{\mu}+igA_{\mu}^a
\frac{\tau^a}{2}+ \frac{ig_1}{2}B_{\mu})L+\nonumber\\
\bar{R}i \gamma_{\mu}(\partial_{\mu}+ig_1B_{\mu})R + \ldots
\label{36}
\end{eqnarray}
where $B_{\mu}$ is the Abelian gauge field, $G_{\mu \nu}$ is the
corresponding stress tensor and $ \ldots$ stands for the similar
terms including quark fields.

The spin zero field Lagrangian is
\begin{eqnarray}
L= |\partial_{\mu}\varphi+ig \frac{\tau^a}{2}A_{\mu}^a \varphi-
\frac{ig_1}{2}B_{\mu} \varphi|^2-G( \bar{L} \varphi R+ \bar{R}
\varphi L)+ \ldots+
\nonumber\\
 \frac{g}{2m_1} \int_{-\pi/\kappa}^{\pi/\kappa} d \lambda
\partial_{\mu}X(x, \lambda)
\partial_{\mu}(\varphi^+ \varphi)+ \frac{1}{2}
[\int_{-\pi/\kappa}^{\pi/\kappa} d \lambda
\partial_{\mu}X(x, \lambda)]^2 -\nonumber\\
\frac{\pi}{\kappa} \int_{-\pi/\kappa}^{\pi/\kappa} d \lambda
\partial_{\mu}X(x, \lambda)
\partial_{\mu}X(x, \lambda)+ \frac{\pi}{\kappa}
\int_{-\pi/\kappa}^{\pi/\kappa} d \lambda a^{-2}( \lambda) \Box
X(x, \lambda) \Box X(x, \lambda) \label{37}
\end{eqnarray}
where again $\ldots$ stand for the terms with quark fields, and
the field $\varphi$ is parameterized as before
\begin{equation}
\varphi_1= \frac{(iB_1+B_2)}{\sqrt{2}}; \quad \varphi_2
=\frac{\sqrt{2}m_1}{g}+ \frac{\sigma-iB_3}{\sqrt{2}} \label{38}
\end{equation}

All the discussion given above is applied to this model. It is gauge
invariant, renormalizable and has no one particle excitations with
spin zero. The interaction of the gauge singlet fields is fixed by
the invariance with respect to the supersymmetry transformations
(\ref{8}).

Notice that the structure of fermion anomalies in our model
remains the same as in the Weinberg-Salam model, so the
compensation of lepton and quark anomalies works in the same way.

\section{Conclusion.}

In this paper we discussed the higher derivative formulation of the
gauge invariant model of the massive Yang-Mills field, which is
unitary in the positive norm space and at the same time does not
produce one particle poles in the channels corresponding to the
exchange by the spin zero particles, and in this sense does not
require the existence of fundamental scalar mesons. A discretized
version describes a gauge invariant model of a massive vector field
with spontaneously broken symmetry and several neutral scalar
mesons. In our model the form of the spin zero field interaction is
fixed by the symmetries imposed on the theory.

This mechanism may be applied to the electroweak model of the
Salam-Weinberg type to modify the predictions concerning the spin
zero particles. The same mechanism may be used in electroweak
models including several fermion generations and in the models
with neutrino oscillations.

For discussion of some experimental consequences of the models with
additional gauge singlet fields see \cite{Kr1}, cite{EsGu},
\cite{Bi} and other papers cited in the introduction.

{\bf Acknowledgements} \\
This work was partially supported by RFBR under grant 050100541,
grant for support of leading scientific schools 20052.2003.1, and
the program "Theoretical problems of mathematics".
 \end{document}